\documentclass{article}
\usepackage{spconf,amsmath,graphicx,makecell,hyperref,textcomp}
\usepackage{enumitem}
\setlist{nosep, leftmargin=14pt}

\usepackage{mwe} 

\begin{document}
\title{Model and predict age and sex in healthy subjects using brain white matter features: A deep learning approach}
%



\name{\makecell{Hao He$^{\star \dagger}$, Fan Zhang$^{\dagger}$, Steve Pieper$^{\dagger}$, Nikos Makris$^{\dagger}$, Yogesh Rathi$^{\dagger}$,\\ William Wells III$^{\dagger}$, Lauren J. O'Donnell$^{\dagger}$}}

\address{$^{\star}$ Glasgow College, University of Electronic Science and Technology of China \\
    $^{\dagger}$ Brigham and Women's Hospital, Harvard Medical School}

\maketitle
\begin{abstract}
The human brain's white matter (WM) structure is of immense interest to the scientific community. Diffusion MRI gives a powerful tool to describe the brain WM structure noninvasively. To potentially enable monitoring of age-related changes and investigation of sex-related brain structure differences on the mapping between the brain connectome and healthy subjects' age and sex, we extract fiber-cluster-based diffusion features and predict sex and age with a novel ensembled neural network classifier. We conduct experiments on the Human Connectome Project (HCP) young adult dataset and show that our model achieves 94.82\% accuracy in sex prediction and 2.51 years MAE in age prediction. We also show that the fractional anisotropy (FA) is the most predictive of sex, while the number of fibers is the most predictive of age and the combination of different features can improve the model performance.
\end{abstract}
\section{Introduction}
\label{sec:intro}

Diffusion magnetic resonance imaging (dMRI) enables in-vivo mapping of the course of white matter (WM) tracts through the brain via a process called tractography \cite{Basser2000-ic}. Many previous works have shown that brain WM tractography contains information that is discriminative of sex and age. For example, sex differences have been identified in intra- vs inter-hemispheric connectivity \cite{Ingalhalikar2014-lj}, and lifespan-related structural connectome changes have been identified in tract microstructure \cite{Lebel2012-mb} and network topology \cite{Zhao2015-jr}. However, the prediction of demographic features such as age and sex using WM tractography is still an interesting and challenging problem. The difficulties are mainly reflected in three aspects. First, it is difficult to choose the combination of diffusion features that is most predictive. Many features can be extracted from dMRI tractography \cite{ZHANG2022}. For example, fractional anisotropy (FA) describes the difference of the tensor ellipsoid's shape from that of a perfect sphere, and mean diffusivity (MD) is a scalar representation of the average of the diffusion tensor's eigenvalues \cite{ODonnell2011-tl}. It is interesting to investigate the diffusion measure or measure combination that best predicts age or sex. Second, it is a technical challenge to extract the features. There are a large number of dMRI preprocessing techniques, which can lead to different processed data and may further result in better or worse prediction performance. Third, the connectome data is high-dimensional. A model that can handle such complex data is needed to learn from extracted connectome features.

Previous studies have focused on both connectome-based sex and age prediction. Some studies describe WM using information from diffusion tensor images (DTI). Mwangi et al. \cite{Mwangi2013-ud} directly trained a relevance vector regression (RVR) classifier to predict subjects' age from four DTI-derived scalar indices (FA/MD/AD/RD). On the other hand, multiple groups have employed structural connectome-based features to describe the WM for the prediction of age and sex. Kulkarni et al. \cite{Kulkarni2013-ny} described the connectome data by using a graph with 70 nodes, where the undirected edge weights represent the number of fibers that pass through any pair of brain regions. Then they train a support vector machine with a non-linear radial basis kernel to perform sex classification. Similarly, Yeung et al. \cite{Yeung2020-ps} encoded the brain connectome into an 85-by-85 adjacency matrix and perform sex prediction using a specialized CNN. To summarize, it is common to conduct machine learning predictions on connectome-based features. But there does not exist a dominant feature description method and learning method which is better than others.
 
In this study, we use a cluster-wise dMRI feature extraction approach that is distinct from those used in previous connectome-based age/sex prediction papers. Whole-brain tractography of each individual is automatically parcellated by a machine learning method based on a neuroanatomist-curated WM atlas \cite{Zhang2018-qx}. Multiple diffusion features are extracted from the tractography parcels. We propose two convolutional neural networks (CNN), a one-dimensional CNN (1D-CNN) and a two-dimensional CNN (2D-CNN), for our prediction tasks. We propose to employ ensemble learning methods to allow different diffusion features to be used simultaneously to improve the network predictivity. Our proposed methods are adaptable to different demographic prediction tasks. In our experiment, sex prediction and age prediction are tested to verify the efficacy of our model.

\section{Methods}
\label{sec:Methods}

\subsection{Feature extraction}
\label{ssec:Feature extraction}

The dMRI data used in this study is part of the Human Connectome Project and was released online on the Connectome Coordination Facility website\footnote{https://www.humanconnectome.org/}. We demonstrate the proposed method using the “1200 Subjects Data Release” dataset version from the HCP young adult study \cite{WU-Minn2017-jg,Van_Essen2013-gd}. The tractography processing pipeline has been discussed in previous work \cite{Zhang2018-qx} and is summarised below.
 
Whole-brain tractography was computed using a two-tensor Unscented Kalman Filter (UKF) method \cite{Reddy2016-wb}, implemented in the ukftractography package via SlicerDMRI \cite{Zhang2020-ui,Norton2017-jt}. The UKF method fits a mixture model of two tensors to the diffusion data while tracking fibers. The two-tensor model associates the first tensor with the main direction of the fiber tract that is being traced, while the second tensor represents fibers crossing through the tract. Whole-brain tractography was visually and quantitatively assessed for quality using the quality control tool in the whitematteranalysis (WMA)\footnote{https://github.com/SlicerDMRI/whitematteranalysis} software. After that, WMA used machine learning to identify WM tracts of each individual based on a neuroanatomist-curated WM atlas \cite{ODonnell2011-tl,Zhang2018-mz,ODonnell2012-kn}. Fiber clusters were classified into three categories based on the location: left-hemisphere/ right-hemisphere/ commissural. Since the atlas clustering is conducted bilaterally, there is a one-to-one mapping between each cluster in the left and right hemispheres. As presented in our experiments, tractography was parcelled into 800 fiber clusters, where 716 clusters are bilateral and the other 84 are commissural. So for each subject, the length of the diffusion features became 716×2+84=1516. Finally, statistical measures were computed within each cluster. Considering that we can compute measurements $\in$(FA/MD) for the tensors $\in$Tensor1/Tensor2) and the statistical values $\in$ (mean/min/max), 12 measures were computed for each cluster. Plus, we also computed the number of fibers/points in each cluster as additional measures. These measures were used as input feature vectors of deep learning models individually or combined. Different statistical measurements may provide complementary information about subjects' sex/age, which is why we try to combine them.

\subsection{Data pre-processing}
\label{ssec:Data pre-processing}

We performed the following pre-processing steps to feed the extracted features into a neural network. First, some clusters could be absent in some subjects due to brain structural differences. Simply removing these clusters would result in inconsistent lengths of input vectors. Instead, the measurement of these missing clusters was set to 0. This could be better than using mean values of other existing cluster measurements since the position of absence may also be a clue for prediction. Then, we rearrange the position of 1516 clusters based on anatomy, grouping clusters from the same  brain regions (right hemisphere, left hemisphere, and commissural brain regions). Last, we conduct a min-max normalization on input vectors after splitting for the cross-validation.

\subsection{1D-CNN and 2D-CNN}
\label{ssec:1D-CNN and 2D-CNN}

For a specific statistical measurement (e.g., the mean FA), the input of the neural network can be represented as either a 1D vector or a 2D matrix. All 1516 elements were arranged as described above in the 1D vector, while elements were arranged based on their anatomical location in the 2D matrix. The shape of the 2D matrix is 3×800, with rows representing the right hemisphere, left hemisphere, and commissural brain regions, respectively, and columns representing 1 to 800 atlas-based cluster numbers. Adapting for the two types of inputs, a one-dimensional (1D) convolutional neural network (CNN) and a two-dimensional (2D) CNN are used and compared. The 1D-CNN takes vectors as input and the 2D-CNN takes a matrix as input. The proposed network structure in this work is adapted from a human activity recognition (HAR) CNN \cite{Zeng2014-xy}.

\begin{figure}[htb]

\begin{minipage}[b]{1.0\linewidth}
  \centering
  \centerline{\includegraphics[width=8.5cm]{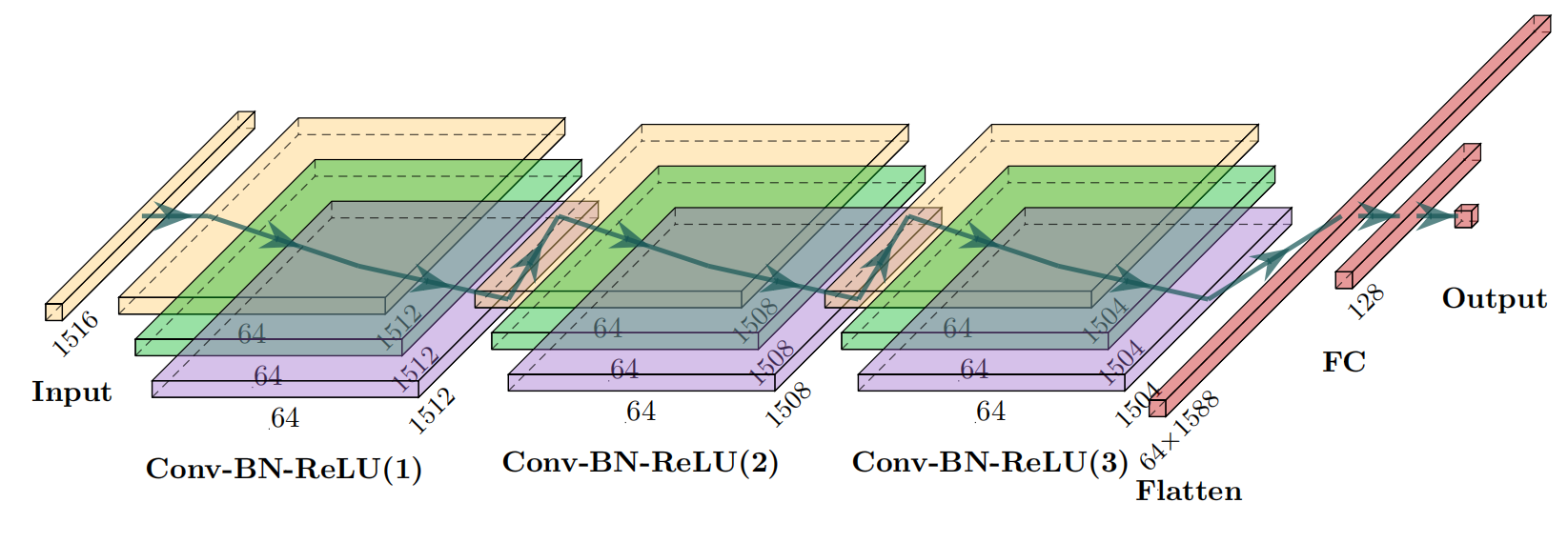}}
  \centerline{(a)}\medskip
\end{minipage}

\begin{minipage}[b]{1.0\linewidth}
  \centering
  \centerline{\includegraphics[width=8.5cm]{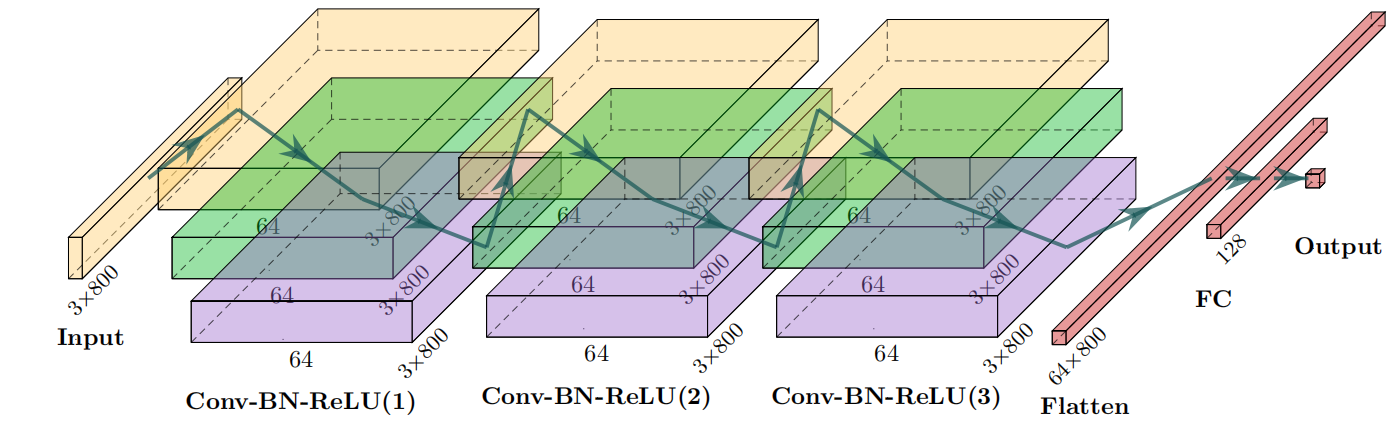}}
  \centerline{(b)}\medskip
\end{minipage}

\caption{ Proposed network structure (a) 1D-CNN, each plane in the Conv-BN-ReLU layer represents a vector with 64 channels and (b) 2D-CNN with input=3×800}
\label{fig:1}
\end{figure}

The 1D-CNN is composed of three convolutional and two fully connected layers. For each convolutional layer, the input vectors go through a 1D convolution layer with kernel size=5, stride=1, filters=64, followed by a batch normalization layer, a ReLU layer, and a dropout layer. Based on our experiments, a stride that is greater than 1 would sharply drop the performance. A possible explanation is that the clusters are relatively independent, so increasing stride will cause information loss from many clusters. Two fully connected layers have 512 and 128 output neurons,  respectively, also followed by a ReLU layer and a dropout layer. This structure is shown in \textbf{Fig.\ref{fig:1}(a)}. As for the 2D-CNN, the main structure remains the same and the convolutional layers are replaced by 2D convolutional layers with kernel size=3×3 and padding=1. At the end of three Conv-layers, a flatten layer was added, converting the 2D matrix into a vector for the fully connected classifier.

\subsection{Ensemble learning}
\label{ssec:Ensemble learning}

The idea of ensemble learning is to build a prediction model by combining the strengths of multiple, simpler base models \cite{Friedman2001-wt}. In our experiments, we choose a method similar to bagging learning, that is, to independently train multiple networks with multiple data splits. One improvement is that the splits are not obtained by bootstrapping the original dataset, but are obtained by using different measurements (described in 2.1). Specifically, weak classifiers, which are mutually independent, are trained from different diffusion measurements. The strong predictions then come from voting for classification tasks and averaging for regression tasks. Ensemble learning methods can reduce model error in two ways. 1) Different measurements can be seen as white matter knowledge from different perspectives. Combining different knowledge, a more comprehensive prediction can be made.  2) Multiple random models can reduce the variance of prediction results, thus increasing the stability of the model.

\section{Experiments and Results}
\label{sec:Experiments and Results}

\subsection{Dataset}
\label{ssec:Dataset}

The original HCP\_1200 dataset contains data from 1206 subjects. Subjects without dMRI data and subjects used for creation of the employed white matter atlas were excluded. In total, 964 subjects were included in this study. Of the 964 subjects, 443 are male (45.6\%) and 521 are female (54.4\%). The age range is 22-37 years old with $\mu$=28.71 and $\sigma$=3.672.

\subsection{Experimental setup}
\label{ssec:Experimental setup}

All experiments were conducted using PyTorch. 5-fold cross-validation was used. In our experiments, batch size was set to 8, dropout rates for all convolutional layers were set to 50\%, and learning rate was set to 0.1. Stochastic Gradient Descent (SGD) optimization algorithm is used to adjust the learning rate in real-time. The training lasted for 300 epochs and the performance of training networks were validated after each epoch. The best testing performance among training epochs was recorded as the final performance of the network. Cross entropy is used for sex classification task and mean square error are used for age prediction.

\subsection{Classifier based on a single feature}
\label{ssec:Classifier based on a single feature}

\begin{table}[h]
\resizebox{86mm}{!}{
		\centering
		\begin{tabular}{cc|cccc}\hline
		    &&\multicolumn{2}{c}{Sex Prediction}&\multicolumn{2}{c}{Age Prediction}\\
		    \hline
			Network&Input&\makecell{Train\\Acc(\%)}&\makecell{Test\\Acc(\%)}&\makecell{Train\\MAE(years)}&\makecell{Test\\MAE(years)}\\
			\hline
			LeNet&FA1-mean&98.19&88.08&2.02&3.8034\\
			2D-CNN&FA1-mean&92.88&92.06&2.02&3.8034\\
			1D-CNN&FA1-mean&95.47&\textbf{92.75}&2.32&\textbf{3.0110}\\
			\hline
		\end{tabular}}
		\caption{Sex/Age Prediction with FA1-mean feature using proposed neural network classifer}
		\label{table:1}
	\end{table}

We first tested the neural network performance with single-feature input to our proposed 1D-CNN/2D-CNN and another well-known CNN, LeNet5. Here FA1-mean (the mean value of tensor one FA, as computed by the UKF method) was used as the input measurement because it is a typical feature that has been studied along fiber tracts \cite{Catani2008-wj}. The result in \textbf{Table \ref{table:1}} shows the 1D-CNN is consistently more predictive both in sex and age prediction, but the 2D-CNN has less overfitting to the dataset (can be inferred from a lower difference between training and testing).

\begin{figure}[htb]

\begin{minipage}[b]{1.0\linewidth}
  \centering
  \centerline{\includegraphics[width=8.5cm]{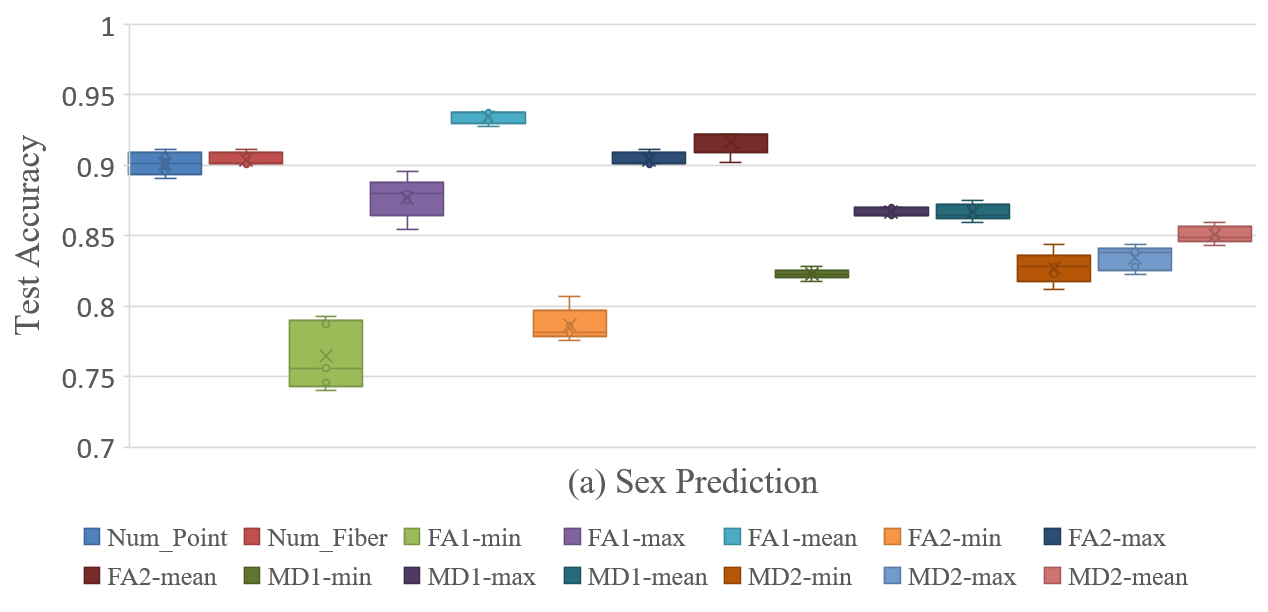}}
\end{minipage}

\begin{minipage}[b]{1.0\linewidth}
  \centering
  \centerline{\includegraphics[width=8.5cm]{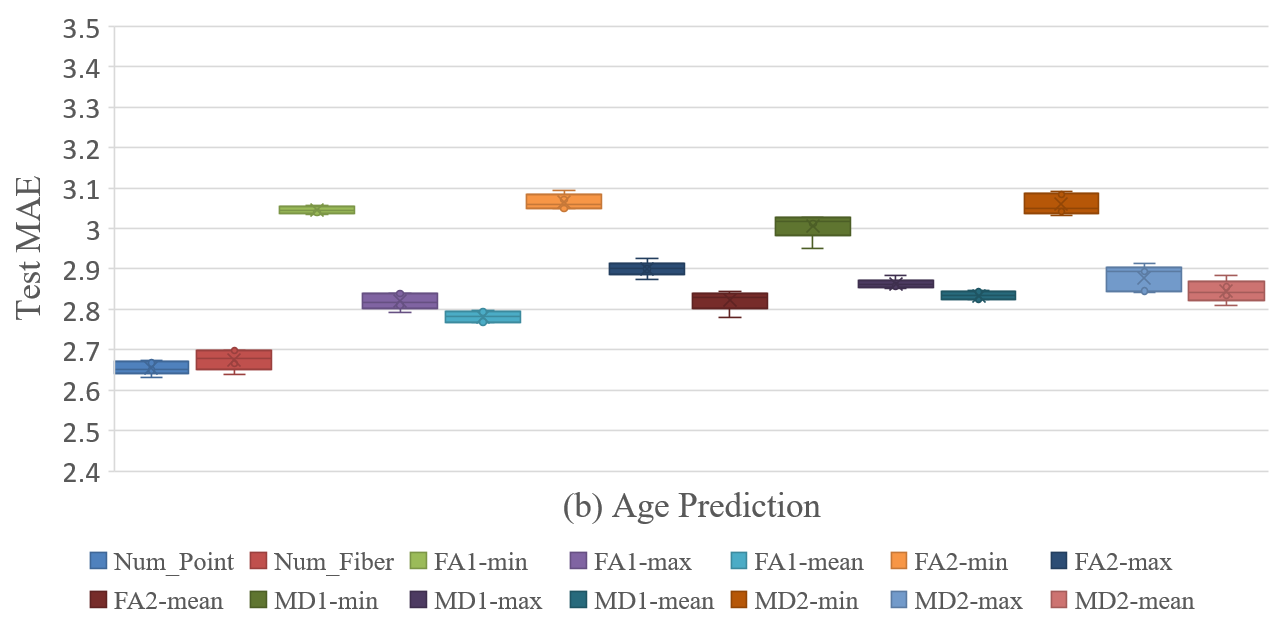}}
\end{minipage}

\caption{box chart of the (a) sex prediction and (b) age prediction performance using 14 different input features with proposed 1D-CNN. Each box contains 5 different data from 5-fold cross-validation.}
\label{fig:2}
\end{figure}

An interesting question is which measurement is more predictive of age/sex. We conducted a comparison experiment using the most predictive model (1D-CNN) verified above to predict both age and sex. 14 types of diffusion features (described in 2.1) were used to train their corresponding networks individually. Based on the experimental result in \textbf{Fig \ref{fig:2}}, training networks with FA/MD/Num\_Fibers/Num\_Points led to quite different performances. FA1-mean is the most predictive feature in sex prediction but number of fiber is the most predictive in age prediction. It can be seen that both in FA/MD, the mean value is more predictive than the max value, and the min value is least predictive.

\subsection{Assembled classifier}
\label{ssec:Assembled classifier}

As discussed before, weak classifiers are assembled to form a stronger classifier. In this experiment, we test the assembled model performance using different feature combinations. Firstly, all chosen features are used to train a 1D-CNN network independently. Then the overall performance of the assembled decision is measured. By comparing it with the single classifier result, we can see clearly the improvement of prediction accuracy.

\begin{table}[h]
\resizebox{86mm}{!}{
		\centering
		\begin{tabular}{cccccccccccccccc}
    		\hline
    		\multicolumn{14}{c}{Input features combination}&\multicolumn{2}{c}{Sex Prediction}\\
    	    \hline
    		1&2&3&4&5&6&7&8&9&10&11&12&13&14&Train acc(\%)&Test acc(\%)\\
    		\hline
    		&&$\surd$&&&&&&&&&&&&98.32&92.75$\pm$0.37\\
    		&$\surd$&$\surd$&&&&&&&&&&$\surd$&&97.93&93.78$\pm$0.45\\
    		&&$\surd$&&&$\surd$&&&$\surd$&&&$\surd$&&&98.83&93.27$\pm$0.49\\
    		&$\surd$&$\surd$&&&$\surd$&$\surd$&&$\surd$&&&$\surd$&$\surd$&&98.19&93.26$\pm$0.20\\
    		&$\surd$&$\surd$&&&$\surd$&&&$\surd$&&&$\surd$&$\surd$&&98.83&\textbf{94.82$\pm$0.21}\\
    		&&$\surd$&&&$\surd$&&&$\surd$&&&$\surd$&$\surd$&$\surd$&99.74&94.27$\pm$0.15\\
    		$\surd$&$\surd$&$\surd$&$\surd$&$\surd$&$\surd$&$\surd$&$\surd$&$\surd$&$\surd$&$\surd$&$\surd$&$\surd$&$\surd$&100.0&94.27$\pm$0.15\\
    		\hline
    		\multicolumn{16}{c}{(a)}\\
    		\hline
    		\multicolumn{14}{c}{Input features combination}&\multicolumn{2}{c}{Age Prediction}\\
    	    \hline
    	    1&2&3&4&5&6&7&8&9&10&11&12&13&14&Train MAE (yrs)&Test MAE (yrs)\\
    		\hline
    		&&$\surd$&&&&&&&&&&&&1.5237&2.768$\pm$0.050\\
    		&&$\surd$&&&$\surd$&&&$\surd$&&&$\surd$&&&1.4519&2.6409$\pm$0.041\\
    		$\surd$&$\surd$&$\surd$&&&$\surd$&&&$\surd$&&&$\surd$&$\surd$&&1.7222&2.585$\pm$0.047\\
    		&&$\surd$&&&$\surd$&&&$\surd$&&&$\surd$&$\surd$&$\surd$&1.2284&\textbf{2.515$\pm$0.036}\\
    		$\surd$&$\surd$&$\surd$&$\surd$&$\surd$&$\surd$&$\surd$&$\surd$&$\surd$&$\surd$&$\surd$&$\surd$&$\surd$&$\surd$&1.0558&2.554$\pm$0.031\\
    		\hline
    		\multicolumn{16}{c}{(b)}\\
		\end{tabular}}
		\caption{Performance of assembled 1D-CNN on (a) sex prediction and (b) age prediction task under different feature combinations. (1:FA1-min, 2:FA1-max, 3:FA1-mean, 4:FA2-min, 5:FA2-max, 6:FA2-mean, 7:MD1-min,  8:MD1-max, 9:MD1-mean, 10:MD2-min, 11:MD2-max, 12:MD2-mean, 13:Num\_Fibers, 14:Num\_Points)}
		\label{table:2}
	\end{table}

\section{Discussion and Conclusions}
\label{sec:Discussion and Conclusions}


In this work, we studied the predictability of individuals' demographic information using diffusion features from brain white matter clusters. The main contributions are as follows. Firstly, we use a novel white matter feature extraction pipeline from dMRI images. Secondly, we propose a deep learning framework to learn from the features, achieving 94.82\% accuracy in sex prediction and 2.51 years MAE in age prediction. Then, we verify by experiments that among dMRI features, FA is more predictive than MD, while mean value is more predictive than max and min value, both in age and sex prediction. Lastly, we verify through experiments that the combination of features improves performance on both sex and age prediction.

Many previous studies have focused on MRI-based sex/age prediction, while relatively few have conducted connectome/dMRI-based prediction. These approaches often represent the features in the form of a DTI image or a structural connectome matrix, followed by a machine learning classifier \cite{Mwangi2013-ud,Kulkarni2013-ny,Yeung2020-ps,Peng2021-tc}. Unlike these approaches, the method proposed in this paper uses a vector of cluster-based diffusion measures to describe the white matter structure. From the literature summarized in \textbf{Table \ref{table:3}}, we can see that deep-learning-based methods have significantly better performance over traditional machine learning. Our method is the SOTA method in connectome-based prediction both in age and sex, but the MRI-based prediction (asterisked in \textbf{Table \ref{table:3}}) still outperforms our method, potentially related to the high amount of information available to the MRI-based method (entire raw T1-weighted images plus various preprocessed inputs such as gray matter segmentations).

\begin{table}[h]
\resizebox{86mm}{!}{
		\centering
		\begin{tabular}{cccccc}\hline
		    Cite&Input feature&\makecell{Train\\Subjects}&\makecell{Learning\\Methods}&\makecell{Sex Pred.\\(accuracy)}&\makecell{Age Pred.\\(MAE)}\\
		    \hline
			\cite{Yeung2020-ps}&$85\times85$ connectome matrix&3152&BrainNetCNN&76.5\%&\\
			\cite{Mwangi2013-ud}&DTI volume (FA)&188&RVR&&5.17\\
			\cite{Kulkarni2013-ny}&70 nodes connectome graph&193&SVM&79\%&\\
			Ours&Cluster-based diffusion measures&964&1D-CNN&94.82$\pm$0.21\%&2.57$\pm$0.04\\
			\cite{Peng2021-tc}$\star$&T1-weighted MRI&1036&SFCN&97.7\%&\\
			\cite{Peng2021-tc}$\star$&T1-weighted MRI&2590&SFCN&&2.76\\
			\hline
		\end{tabular}}
		\caption{Summarization of studies in the field of connectome-based sex/age prediction, as well as T1w-MRI-based predictions (asterisked) for comparison.}
		\label{table:3}
	\end{table}

Besides the performance, there are some other interesting findings in our work. Firstly, inferring from the result in \textbf{Fig \ref{fig:2}}, FA is more predictive of sex than MD, number of fibers, or Num\_Points, while number of fibers is the most predictive of age. For the investigated prediction tasks, the mean value is always more predictive than the max value, and the min value is far less discriminative than the mean and max. Secondly, simply combining more measures does not always improve the prediction performance. Instead, we should select a subset of features for best performance (\textbf{Table \ref{table:2}}). Finally, this work shows the potential of combining diffusion features for the study of the healthy brain's structural connectome.

One limitation of this work is that only one dataset (HCP Young Adult) was used and therefore the age range was limited (22-37 years old). A future improvement of this work could be the use of more general datasets to verify the validity of the method across all age groups.

\section{Compliance with Ethical Standards}
\label{sec:Compliance with Ethical Standards}

This research study was conducted retrospectively using human subject data made available in open access by  HCP \cite{Van_Essen2013-gd}. Ethical approval was not required.

\section{Acknowledgments}
\label{sec:acknowledgments}

We acknowledge the following NIH grants: P41EB015902, R01MH074794, R01MH125860 and R01MH119222. F.Z. also acknowledges a BWH Radiology Research Pilot Grant Award.

\small
\bibliographystyle{IEEEbib}
\bibliography{refs}

\end{document}